# Enhanced Energy-Aware Feedback Scheduling of Embedded Control Systems

Feng Xia[1,4], Longhua Ma[2], Wenhong Zhao[3], Youxian Sun[2], Jinxiang Dong[1]

[1]College of Computer Science and Technology, Zhejiang University, Hangzhou 310027, China
Email: f.xia@ieee.org

[2]State Key Lab of Industrial Control Technology, Zhejiang University, Hangzhou 310027, China
Email: lhma@iipc.zju.edu.cn

[3]College of Mechanical and Electrical Engineering, Zhejiang University of Technology, Hangzhou 310032, China

[4]Faculty of Information Technology, Queensland University of Technology, Brisbane QLD 4001, Australia

*Abstract*— Dynamic voltage scaling (DVS) is one of the most effective techniques for reducing energy consumption in embedded and real-time systems. However, traditional DVS algorithms have inherent limitations on their capability in energy saving since they rarely take into account the actual application requirements and often exploit fixed timing constraints of real-time tasks. Taking advantage of application adaptation, an enhanced energy-aware feedback scheduling (EEAFS) scheme is proposed, which integrates feedback scheduling with DVS. To achieve further reduction in energy consumption over pure DVS while not jeopardizing the quality of control, the sampling period of each control loop is adapted to its actual control performance, thus exploring flexible timing constraints on control tasks. Extensive simulation results are given to demonstrate the effectiveness of EEAFS under different scenarios. Compared with the optimal pure DVS scheme, EEAFS saves much more energy while yielding comparable control performance.

## I. INTRODUCTION

Power management has become a critical design issue, particularly in battery operated real-time embedded systems. Low power design not only reduces the operational cost but also increases the system reliability, while prolonging the battery's lifetime [1]. Dynamic voltage scaling (DVS) [2,3] is one of the most effective approaches to power consumption reduction. However, conventional real-time DVS algorithms rarely take into account the resulting performance of target applications when determining the voltage level of the processor. Though much effort has been made on DVS for real-time applications, e.g. [2,4,5], state-of-the-art DVS algorithms usually rely on fixed timing constraints of real-time tasks. They typically derive the processor speed that provides timeliness guarantees during runtime according to pre-specified periods/deadlines of the task set, and these timing attributes will never be intentionally changed in favour of energy savings, e.g., in response to the actual application requirements.

In practice, however, the resources that an application demands may vary over time. One representative example is control systems. From the control perspective, smaller sampling periods are beneficial to rapid recovery of steady states. Consequently, the negative effect of perturbations will be alleviated, and the quality-of-control (QoC) will then be improved. When the system is in a steady state, however, an unnecessarily small sampling period implies waste of resources (e.g., CPU time and energy). In this case, the sampling period may be enlarged to some extent without significantly degrading the control performance [6-10]. This feature of real-time control applications makes it possible to dynamically allocate CPU resource to each control task according to their real demands.

Improving QoC and reducing energy consumption pose conflicting requirements. The objective of this paper is to develop an approach to reduce CPU energy consumption while preserving QoC guarantees. The effects of sampling periods on energy consumption and QoC will be exposed via motivating examples, respectively. An enhanced energy-aware feedback scheduling (EEAFS) scheme will be proposed, which takes advantage of application adaptation. In particular, the proposed scheme has the following features:

- It integrates feedback scheduling with DVS, providing an effective way for managing QoC and energy consumption simultaneously in embedded real-time control systems. DVS provides an enabling technology for feedback schedulers to manipulate the tasks' execution times, while feedback scheduling enables further energy savings over pure DVS schemes.
- By exploiting direct feedback scheduling [11,12], the sampling periods of control loops (in addition to the CPU speed) are adjusted dynamically to make better compromise between application performance and energy consumption. In other words, the proposed scheme utilizes flexible timing constraints of real-time tasks to enhance the performance of DVS in saving energy. This is in contrast to most previous DVS algorithms that feature only adaptation of CPU speed.
- Task execution times will be indirectly altered at runtime because of the adaptation of CPU speed via DVS. Therefore, the proposed scheme is a new type of feedback scheduling schemes that adapts multiple timing attributes of real-time tasks simultaneously, i.e., it alters both periods and execution times of the real-time control tasks.

Limited work has been found in the literature on DVS for real-time control systems. Solutions for integrated optimization of sampling periods and CPU speed have been presented in [13,14]. Jin *et al.* [15] developed a feedback fuzzy-DVS scheduling method. Marinoni and Buttazzo [16] presented a method that integrates elastic scheduling with DVS management to fully exploit the available computational resources in processors with limited voltage levels. In our previous work [17], a control theoretic approach to DVS for embedded control systems was explored. We have also developed a simple yet efficient DVS scheme that combines time-triggered and event-triggered mechanisms [18]. In contrast to all these reports, the focus of this paper is on exploiting graceful degradation of application performance within the framework of direct feedback scheduling to achieve further energy consumption reduction over pure DVS schemes. The proposed approach extends our previous work [19,20], in which the preliminary idea of enhancing DVS with application adaptation was presented with only limited results.

Feedback scheduling [10,11] has recently emerged as a promising technology for resource management. In particular, significant effort has been made on feedback scheduling of control systems, e.g., [7-9,12]. But none of them deal with power management. There are also several papers applying feedback control technology to DVS. For instance, the popular PID (Proportional-Integral-Derivative) control has been integrated into different DVS algorithms [5,21,22]. However, these algorithms have been used for general-purpose computing systems other than control applications.

The rest of this paper is structured as follows. In Section II the model of the system under consideration is characterized, and an optimal pure DVS scheme is described, which serves as both a building block for the proposed approach and the baseline for comparison. Section III examines the impact of sampling periods on energy consumption and QoC, respectively. The motivation of this work is illustrated by means of case studies. Section IV describes the architecture of EEAFS, two alternative feedback scheduling algorithms, and the theoretical analysis of EEAFS' performance. In Section V, several sets of simulation experiments are carried out assessing the performance of the proposed approach. Finally, Section VI concludes the paper.

## II. OPTIMAL PURE DYNAMIC VOLTAGE SCALING

Consider a DVS-enabled multitasking embedded processor, which is responsible for executing $N$ independent control tasks. Each control task corresponds to a physical process. The supply voltage and operating frequency (CPU speed) can be adjusted with a scaling factor $\alpha \in [\alpha_{\min}, 1]$. Hereafter $\alpha$ is also used to denote the (normalized) CPU speed. Assume that the CPU speed can change continuously in the range of $[\alpha_{\min}, 1]$. The major timing attributes of a control task $i$ are described below.

- $h_i$: period of task $i$, equal to the sampling period, with a nominal (initial) value of $h_{i,0}$.
- $c_{i,nom}$: execution time of task $i$ at full CPU speed, i.e. when $\alpha = 1$.
- $c_i$: actual execution time of task $i$ when CPU speed is scaled, and it holds that $c_i = c_{i,nom}/\alpha$.

All the above timing parameters are variable yet available at runtime. By default, the relative deadline of a control task equals its period under all circumstances. CPU *utilization* $U = \sum c_i / h_i$, while CPU *workload* $\omega = U \cdot \alpha = \sum c_{i,nom} / h_i$.

The system utilizes the earliest deadline first (EDF) algorithm as the underlying scheduling policy. Accordingly, the upper bound of schedulable CPU utilization is 100%, and the schedulability condition is:

$$\sum_{i=1}^{N} \frac{c_i}{h_i} \leq 1 \Leftrightarrow \omega \leq \alpha . \quad (1)$$

Assume that: 1) $\sum_{i=1}^{N} \frac{c_{i,nom}}{h_{i,0}} \leq 1$, and 2) the switching overheads between different voltage levels of the processor is negligible. In reality, switching between different voltage levels takes time and consumes energy. However, in most cases the switching time of prevailing processors is negligibly small in comparison with control task periods. The (normalized) energy consumption of the processor is calculated as [4]:

$$E(\alpha) = \alpha^2 . \quad (2)$$

For the above system, the following theorem, which is derived from the work by Sinha and Chandrakasan [4], gives the optimal voltage level $\alpha_{opt}$.

**Theorem 1.** For an embedded system encompassing a set of periodic control tasks characterized by the timing attributes described above, the processor will consume the minimum energy while meeting task schedulability constraint if and only if the CPU speed is set to:

$$\alpha_{opt} = \sum_{i=1}^{N} \frac{c_{i,nom}}{h_i} = \omega . \quad (3)$$

A general understanding of Theorem 1 is that if the supply voltage level is set to $\alpha_{opt}$, the CPU energy consumption will be the minimum in a sense that the CPU utilization will be maximized while the task set is schedulable with EDF. For a pure DVS scheme, it is the best way to scale voltage according to (3). Therefore, this scheme is called the optimal pure DVS, abbreviated to opDVS. This paper will construct EEAFS from opDVS, and compare their performance. For simple description, assume that $\alpha_{\min}$ is small enough such that $\alpha_{\min} \leq \omega$.

In real processors the speed of instruction execution may not be strictly proportional to clock frequency. It is also an approximately (not strictly) proportional relationship between energy consumption and square of a speed scaling factor. However, these will not affect the applicability of EEAFS because it does not depend on how the task execution times and the energy consumption changes with CPU speed. Given that the DVS technology is supported, EEAFS can be applied to achieve additional energy saving over pure DVS schemes.

### III. MOTIVATING EXAMPLES

#### A. Energy Consumption with Different Sampling Periods

Examined first is how the sampling periods of control loops (i.e. task periods) impact the energy consumption of the processor. It follows from (2) and (3) that

$$E(\alpha) = \alpha_{opt}^2 = \left( \sum_{i=1}^{N} \frac{c_{i,nom}}{h_i} \right)^2. \quad (4)$$

Given below is an illustrative example.

**Example 1.** Consider a system with two control tasks. It is known that $c_{1,min} = 4$, $c_{2,min} = 5$, $h_1 \in [10, 20]$, $h_2 \in [10, 30]$. All these parameters are given in the same time unit. When opDVS is employed, the CPU energy consumption as a function of $h_1$ and $h_2$ is depicted in Figure 1.

It is seen from Figure 1 that the CPU energy consumption decreases with increasing sampling periods. Therefore, increasing sampling periods benefits energy saving.

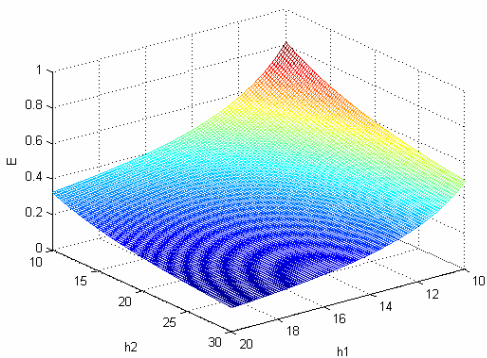

Figure 1. Normalized CPU energy consumption in Example 1

#### B. Quality of Control with Variable Sampling Periods

As pointed out above, it is possible to reduce energy expenditure by using larger sampling periods. However, according to sampled-data control theory, an increase in sampling period degrades control performance. Therefore, with the conventional framework of fixed timing constraints, the effectiveness of saving energy by increasing sampling periods is often impaired by significant QoC deterioration, which is not allowed in most cases.

To address this problem, we attempt to impose flexible timing constraints on control tasks. The following example is employed to illustrate the motivation and to demonstrate the feasibility of this idea. Note that this example is not for describing the EEAFS scheme or evaluating its performance, which will be done in Sections IV and V, respectively.

**Example 2.** Consider a control loop in which a DC motor modelled by $G(s) = 1000/(s^2+s)$ is controlled using PID algorithm. The controller parameters are well designed and remain the same as those in [23], with an initial sampling period of 6ms. The control task runs on a dedicated processor. Assess the system performance in the following two cases:

- Case I: The sampling period remains constant during runtime.
- Case II: Increase the sampling period from 6 ms to 12 ms when the system is in steady state.

Figure 2 shows the system responses and task executions under both cases.

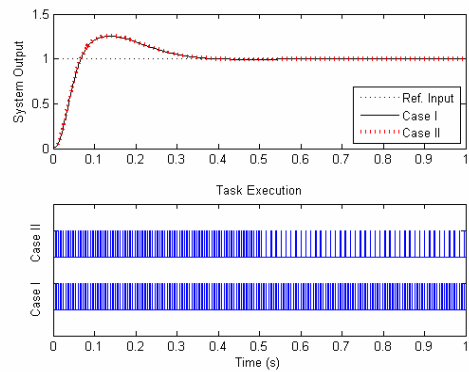

Figure 2. System performance in Example 2

From the upper part of Figure 2, it can be found that in Example 2 the same QoC is achieved under two different cases. In the second case, though the sampling period increases after time instant t = 0.5s (i.e. when the system is in steady state), the resulting control performance does not deteriorate. On the other hand, the lower part of Fig 2 indicates that the increase in sampling period remarkably reduces the consumption of the computing resources. In this example, the CPU time that the control task demands decreases to half of the original after t = 0.5s. According to the discussion in Section III, it is believed that the CPU energy consumption could be significantly cut.

It can be outlined from the above analysis that the sampling period can be properly increased without jeopardizing QoC when the relevant control error is relatively small. This methodology, also referred to as *graceful gradation,* benefits energy saving and does not significantly deteriorate control performance.

### IV. ENHANCED ENERGY-AWARE FEEDBACK SCHEDULING

Based on the above observations, an enhanced energy-aware feedback scheduling scheme will be proposed in this section. Within the same framework, two alternative algorithms are given for sampling period adjustment.

#### A. Framework

The framework of the proposed scheme is shown in Figure 3. A time-triggered feedback scheduler, with an invocation interval of $T_{FS}$ is utilized. For simplicity, assume that the execution times (at full CPU speed) of all tasks would never change except for at invocation instants of the feedback scheduler. Actually, this assumption is not imperative for EEAFS since it is applicable provided that the execution times are available, even when they are time-varying. In this context, the execution time and period of each task are deterministic in every invocation interval, which makes CPU utilization also known. As a consequence, the actual CPU utilization will be kept at the desired level each time the feedback scheduler is executed. In other words, if EEAFS is used, the resulting CPU utilization will always be kept at the desired level, 100% for example in this paper, assuming that the time overhead of the feedback scheduler is negligible. Therefore, in the framework of the EEAFS, there is actually no real feedback information about CPU utilization. This is primarily due to the availability of accurate timing attributes of tasks.

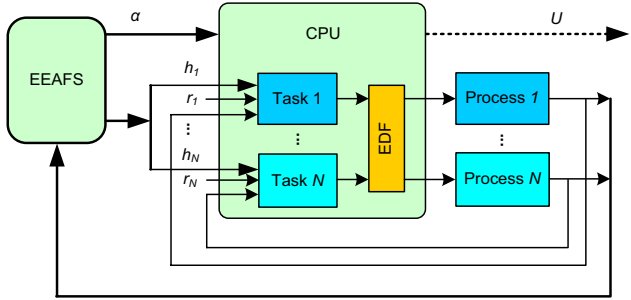

Figure 3. Framework of enhanced energy-aware feedback scheduling

With the EEAFS, the current absolute control error of each control loop will be fed back to the feedback scheduler at every invocation instant. The *absolute control error* $e_i$ is defined as the absolute difference between the reference input $r_i$ and the system output $y_i$, i.e., $e_i = |r_i - y_i|$.

The feedback scheduler consists of two main parts: sampling period adjustment and voltage scaling. The former is responsible for adjusting the sampling period of each loop based on the feedback about its current control error. The relevant algorithm will be given in the next subsection. The latter exploits the opDVS scheme described in Section II. Its role is to scale the voltage/frequency of the processor using the DVS technique. The adoption of the opDVS implies setting the desired CPU utilization level to the system schedulability bound (i.e., 100%). This gives rise to the maximum utilization of CPU resources and hence the lowest possible energy expenditure.

In the above description, it is assumed that the sampling periods of all control loops remain constant in the course of every invocation interval. In practice, it is possible that the following situation happens. At the $j$-th invocation instant (of the feedback scheduler), the sampling period of a loop, say loop $i$, is set to the maximum $h_{i,max}$ because the controlled process is in a steady state. At a certain time instant $t$ ($jT_{FS} < t < (j+1)T_{FS}$), process $i$ is disturbed by a sharp perturbation, e.g. a step change input. Since the sampling period at this time remains to be $h_{i,max}$, which is too large for the controller to response quickly enough, the corresponding control performance may be significantly degraded.

To deal with this problem, an event-triggered mechanism is introduced for the feedback scheduler in addition to the time-triggered activation. The basic idea behind is that if only the absolute change of control error in a control loop exceeds a specific threshold, the feedback scheduler will be invoked immediately. Accordingly, the condition of producing an event that triggers the feedback scheduler is given by:

$$\Delta e_i(t) = |e_i(t) - e_i(jT_{FS})| > \delta \qquad (5)$$

where $\delta$ is a design parameter denoting the threshold of $\Delta e_i$. It is possible to choose different $\delta$ values for different control loops. When the feedback scheduler is triggered by an event rather than a timer, it will only re-assign the sampling period of the control loop that generates the event, while others remain unchanged. The CPU speed will still be re-set according to the opDVS. One of the major reasons for this operation is to cut down the scheduling overhead.

The combination of traditional time-triggered mechanism and the event-triggered mechanism makes it much easier to choose an appropriate invocation interval $T_{FS}$. In practice, this choice can be made based on related characteristics of the system, such as the frequency of variations in task execution times, the magnitude and frequency of perturbations, etc. If $T_{FS}$ is a little bigger than necessary, the even-triggered mechanism will compensate for its negative effect, at least in part.

### B. Period Adjustment Algorithms

A key issue in EEAFS design is the algorithm for sampling period adjustment. The basic principle of the sampling period adjustment is as follows. When the current QoC is good, the corresponding control task will be assigned large periods. Conversely, small periods will be assigned in the case of bad QoC. Provided that the schedulability constraint is preserved, the calculation of sampling periods will be independent from one loop to another. For simple notation, the subscript $i$ will hereafter be dropped from all variables wherever possible.

A prerequisite for online assignment of sampling periods is to select a proper metric to indicate instantaneous control performance. A reasonable choice is the absolute control error. It is generally acknowledged that bigger errors indicate worse QoC. However, if the process output sharply oscillates around the reference input, the absolute control error might still be small sometimes, which obviously does not reflect the real control performance. To address this problem, the following instantaneous control performance index, denoted *ind*, is defined:

$$ind(j) = \lambda \cdot ind(j-1) + (1-\lambda) \cdot e(j) \qquad (6)$$

where $\lambda$ is a forgetting factor.

To adapt the sampling period $h$, it is necessary to determine its allowable range first. In general, the maximum allowable value of $h$ can be obtained from the stability condition of the control system. In this paper, the maximum sampling period $h_{max}$ is instead determined by simulations. This reduces the dependency of EEAFS on the system models of control loops while making it more practical. Still, the theoretical results on $h_{max}$ will serve as a useful reference.

With regard to the minimum allowable value of $h$, it is set for simplicity that $h_{min} = h_0$, i.e., the minimum allowable period is equal to its nominal value. As a matter of fact, it is possible in most cases that a sampling period less than $h_0$ is assigned when the control loop encounters a severe perturbation such that the control performance is further improved without violating the system schedulability. However, this will adversely complicate the problem. If sampling periods less than $h_0$ are allowed, much care should be put on the system schedulability. On the other hand, since it is assumed that $\sum_{i=1}^{N} \frac{c_{i,nom}}{h_{i,0}} \leq 1$, the runtime workload is given by

$$\omega = \sum_{i=1}^{N} \frac{c_{i,nom}}{h_i} \leq \sum_{i=1}^{N} \frac{c_{i,nom}}{h_{i,0}} \leq 1 \text{, if } h_{min} = h_0. \text{ That is, the}$$

feasibility of the voltage scaling module within the EEAFS is then always guaranteed.

Based on the above discussions, the allowable range of sampling periods can be specified accordingly. Described next is how to determine the sampling period dynamically. The equation used here to calculate the new period is:

$$h(j) = \eta(j) \cdot h_{min} = \eta(j) \cdot h_0 \qquad (7)$$

where $\eta$ is *period scaling factor*, and satisfies $1 \leq \eta \leq h_{max}/h_{min}$. Since the major role of EEAFS is to minimize energy consumption without jeopardizing QoC, $\eta$ needs to be maximized as much as possbile. In the following, two alternative algorithms are presented to compute $\eta$.

*1) Exponential Algorithm*

In this case $\eta$ is computed as:

$$\eta(j) = \begin{cases} h_{max}/h_{min} & \text{if } ind(j) \leq e_{min} \\ \frac{e^{-\beta \cdot ind(j)} - e^{-\beta \cdot e_{max}}}{e^{-\beta \cdot e_{min}} - e^{-\beta \cdot e_{max}}}(h_{max}/h_{min} - 1) + 1 & \\ & \text{if } e_{min} < ind(j) < e_{max} \\ 1 & \text{if } ind(j) \geq e_{max} \end{cases} \qquad (8)$$

where $\beta$ is a constant introduced to enhance the effect of the exponential function, $e_{min}$ and $e_{max}$ are design parameters that decide the range of the instantaneous control performance index in which sampling period can be adjusted arbitrarily.

According to (7) and (8), the sampling period will be directly set to be the minimum, which corresponds to $\eta = 1$, once the performance index $ind(j)$ exceeds a upper threshold $e_{max}$. The goal of this operation is to improve control performance via quick response to large derivations. In contrast, the period will be the maximum, i.e. $\eta = h_{max}/h_{min}$, if $ind(j)$ becomes less than another limit $e_{min}$, which implies that the system approaches a steady state. The maximum period is set to achieve the largest possible energy saving. In other cases, i.e., when $e_{min} < ind(j) < e_{max}$, a period that decreases exponentially with increasing $ind(j)$ will be assigned. With the exponential function, the period will rapidly decrease as the performance index (control error) increases. This benefits the improvement of control performance. As $\beta$ increases, the effect of the exponential function will be enhanced. Since a small $\eta$ value yields a relatively small sampling period, the algorithm becomes more aggressive with smaller $\beta$ in view of energy saving.

*2) Linear Algorithm*

A more straightforward method to adjust sampling periods is the linear function given below.

$$\eta(j) = \begin{cases} h_{max}/h_{min} & \text{if } ind(j) \leq e_{min} \\ h_{max}/h_{min} - \frac{ind(j) - e_{min}}{e_{max} - e_{min}}(h_{max}/h_{min} - 1) & \\ & \text{if } e_{min} < ind(j) < e_{max} \\ 1 & \text{if } ind(j) \geq e_{max} \end{cases} \qquad (9)$$

When $ind(j) \geq e_{max}$ or $ind(j) \leq e_{min}$, the computation of this linear function is the same as the exponential function. Nevertheless, they are different when $e_{min} < ind(j) < e_{max}$. In these cases, the linear algorithm adjusts sampling periods linearly. Its curve is exactly a line that goes across points $(e_{min}, h_{max}/h_{min})$ and $(e_{max}, 1)$ on the plane. Compared with the exponential function, the linear function is simpler and straightforward, but less flexible.

In both algorithms, it is essential to choose proper values for $e_{min}$ and $e_{max}$. For the purpose of improving QoC, $e_{min}$ and $e_{max}$ should be minimized. However, from the perspective of energy saving, large $e_{min}$ and $e_{max}$ should be chosen. The principle for choosing $e_{min}$ and $e_{max}$ is to use as large values as possible given that the QoC is not jeopardized. This facilitates further energy saving and hence improves the energy efficiency of the system.

*C. Analysis*

The proposed scheme can be described in Figure 4. Both the CPU speed and the periods of the control tasks are adapted with this scheme. Since a lower CPU speed yields larger task execution times, reducing CPU speed and decreasing periods conflict with each other in the presence of resource schedulability constraints. Both of them cause the requested CPU utilization of control tasks to increase. Because EEAFS enlarges sampling periods on condition that the control performance is not significantly degraded, it is capable of saving additional energy while providing comparable QoC.

When the EEAFS scheme is applied, the maximum and minimum possible values of normalized CPU energy consumption are given by the following theorem.

**Theorem 2.** For the multitasking embedded control system described in Section II, if the algorithm depicted by Figure 4 is employed, then the range of normalized CPU energy consumption is $[E_{min}, E_{max}]$, where

$$E_{min} = \left(\sum_{i=1}^{N} \frac{c_{i,nom}}{h_{i,max}}\right)^2, \quad E_{max} = \left(\sum_{i=1}^{N} \frac{c_{i,nom}}{h_{i,0}}\right)^2 \qquad (10)$$

```
//h: Sampling period
//α: Normalized CPU speed
Enhanced Energy Aware Feedback Scheduling {
    IF triggered by timer
        FOR each control loop
            Sample the process output y;
            e←abs(r-y);
            //r: reference input
            Compute ind using (6);
            Compute η using (8) or (9);
            //η: period rescaling factor
            Compute h using (7);
            //Update task period;
        END
    ELSE (triggered by loop i)
        Update h_i using (6)-(9);
    END
    Compute α using (3);
    Assign CPU speed to α;
}
```

Figure 4. Pseudo code for enhanced energy-aware feedback scheduling

The proof of this theorem is straightforward. Some brief arguments are given below.

Equation (4) indicates that the processor will consume the minimum energy if all task periods take on their maximum possible values. According to the algorithms given in Figure 4, the maximum possible value of $h_i$ is $h_{i,max}$. If $ind(j) \leq e_{min}$ is satisfied simultaneously for all control loops, the CPU energy consumption reaches then the minimum, and it holds that $E = E_{min} = \left(\sum_{i=1}^{N} \frac{c_{i,nom}}{h_{i,max}}\right)^2$.

Similarly, energy consumption reaches its maximum value $E = E_{max} = \left(\sum_{i=1}^{N} \frac{c_{i,nom}}{h_{i,0}}\right)^2$ when $ind(j) \geq e_{max}$ is satisfied for all loops. In Example 1, for instance, the minimum possible (normalized) energy consumption is $E_{min} = (4/20+5/30)^2 \times 100\% = 13\%$, while the maximum value is $E_{max} = (4/10+5/10)^2 \times 100\% = 81\%$.

## V. PERFORMANCE EVALUATION

Several sets of comparative simulations are conducted using Matlab with TrueTime [23]. Consider an embedded control system with four independent control loops. The controllers are of PID type and are well tuned. Digital controllers are designed by discretizing continuous-time controllers [9], which can be described by $G_{PID}(s) = K_P + \frac{K_I}{s} + K_D s$. The system models of the controlled processes and the corresponding controller parameters are given in Table I, where the time unit of task execution times and periods is $ms$. These four control tasks, along with the feedback scheduling task, run concurrently on one processor. The overhead of executing the feedback scheduler is neglected. For the purpose of simple description, the $c_{nom}$ values of all tasks are fixed in the simulations. Because all timing attributes of the tasks are precisely known, the proposed scheme will perform in like manner in the case of variable $c_{nom}$ values. In addition, the following parameters are used in all simulation experiments: $T_{FS}$ = 50 ms, $\lambda$ = 0.3, $e_{max}$ = 0.2, and $e_{min}$ = 0.02. All perturbations on control loops are step input changes.

TABLE I.
SETUP OF SIMULATED SYSTEM

| | System Model | Controller Parameters | $c_{nom}$ | $h_0(h_{min})$ | $h_{max}$ |
|---|---|---|---|---|---|
| Loop 1 | $\frac{1}{1000s+50}$ | $K_P=10^4$, $K_I=400$, $K_D=0$ | 2 | 10 | 40 |
| Loop 2 | $\frac{1}{s^2+10s+20}$ | $K_P=30$, $K_I=70$, $K_D=0$ | 2 | 7 | 30 |
| Loop 3 | $\frac{1}{0.5s^2+6s+10}$ | $K_P=100$, $K_I=200$, $K_D=2$ | 2 | 8 | 30 |
| Loop 4 | $\frac{1}{s^2+10s+20}$ | $K_P=200$, $K_I=350$, $K_D=3$ | 2 | 9 | 40 |

To measure the QoC quantitatively, the Integral of Absolute Error (IAE) is recorded respectively for each loop, i.e., $J_i(t) = \int_0^t e_i(\tau)d\tau$. The total control cost of the system is calculated as $J_{SUM}(t) = \sum_{i=1}^{4} J_i(t)$. Normalized CPU energy consumption is computed using (2).

### A. Different Schemes

In this set of simulations, the following schemes are compared:

- opDVS: the optimal pure DVS scheme with fixed task period equal to the nominal sampling period.
- EEAFS-1: the proposed EEAFS scheme that adopts the exponential period scaling algorithm, with $\beta$ = 40.
- EEAFS-2: the proposed EEAFS scheme that adopts the linear period scaling algorithm.

The simulation runs as follows. At time t = 0, loop 1 and loop 2 start running, and loop 1 is disturbed by a step input change. Loop 2 is also perturbed at time t = 2s. Loops 3 and 4 remain off until t = 4s, at which both of them encounter step change input. At time instant t = 6s, all loops are perturbed simultaneously. The whole run ends at t = 8s.

*1) Energy Consumption*

Figure 5 shows the normalized CPU energy consumption under the different schemes. As an optimal pure DVS scheme, the opDVS is effective in energy consumption reduction, especially when the workload is light, e.g., in the time interval t = 0-4s. Compared to the opDVS, both EEAFS-1 and EEAFS-2 are able to save much more energy. Throughout the simulations, the average energy consumption under three schemes is 57.9%, 10.9%, and 9.7%, respectively. The CPU energy consumption under EEAFS-1 and EEAFS-2 decreases over opDVS by 47.0% and 48.2%, respectively, on average. It is also found that the CPU utilization remains at 100% in all simulations. This implies that CPU time is fully utilized under all three schemes.

Each task's period is depicted in Figure 6. It is seen that: 1) the sampling periods are the largest under EEAFS-2 almost all the time, 2) the sampling periods under EEAFS-1 are slightly smaller than, but very close to, those under EEAFS-2, and 3) the sampling periods under the opDVS are always the shortest and fixed.

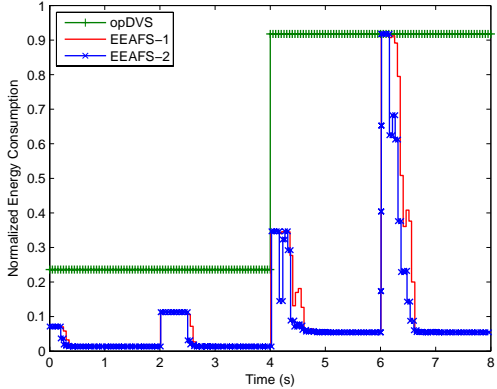

Figure 5. Normalized energy consumption under different schemes

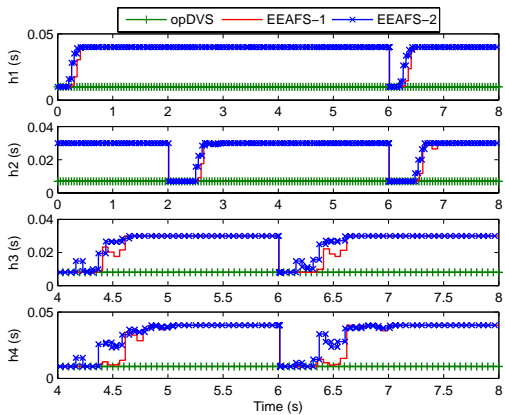

Figure 6. Task periods under different schemes

Due to the fact that the task periods are enlarged, EEAFS-1 and EEAFS-2 are capable of saving more energy over the opDVS. For instance, in time intervals t = 5-6s and 7-8s, all task periods are set to the maximum possible values, i.e. $h_i = h_{i,max}$ when EEAFS is applied. This results in a low energy consumption of 5.5%, which is 86.3% less than that of the opDVS case.

*2) Quality of Control*

Figure 7 depicts the total control cost $J_{SUM}$ of the system. It is seen that all three schemes deliver comparable overall control performance. The accumulative total control costs are 1.309, 1.396, and 1.443, respectively. Compared with the opDVS case, the total control costs under EEAFS-1 and EEAFS-2 increase by 6.7% and 10.2%, respectively.

To summarize this set of experiments, both the exponential and linear EEAFS schemes can achieve significant additional energy consumption reduction over the optimal pure DVS scheme, while delivering comparable control performance. The linear EEAFS algorithm is more aggressive in energy saving than the exponential EEAFS algorithm (with $β = 40$), but with slightly degraded QoC. Overall, these two algorithms perform comparably. In the following, unless otherwise specified, only the exponential EEAFS algorithm will be studied thanks to its higher flexibility in design.

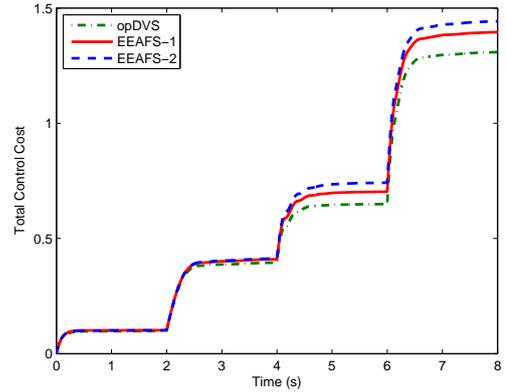

Figure 7. Total control costs under different schemes

*B. Different Design Parameters*

Next the performance of the EEAFS with different design parameters is assessed. In particular, the effect of different $β$ values on EEAFS is studied. The simulation pattern remains the same as those described in Section 5.1. The set of $β$ values chosen for simulations is {1, 10, 20, 40, 60, 80, ∞}, where EEAFS with $β→∞$ implies a scheme similar to the dynamic solution in [13].

Figure 8 shows the average (normalized) energy consumption for different $β$ values. It is observed that the energy expenditure increases with the increase in $β$ value, implying that smaller $β$ values are more beneficial to energy saving. Even in the worst case where $β→∞$, however, the EEAFS is still able to reduce 41.7% additional energy consumption on average over opDVS.

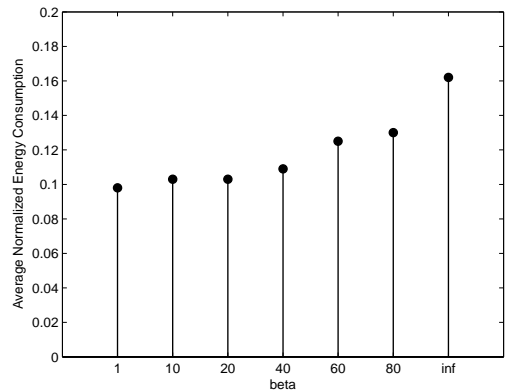

Figure 8. Normalized energy consumption with different parameters

As Figure 9 shows, for each control loop, the difference between accumulated control costs with different $β$ values is minor. The overall control performance under the EEAFS with different $β$ values is quite comparable. In general, decreasing $β$ will reduce energy expenditure, but may possibly cause slight control performance degradation. Indeed, the choice of $β$ corresponds to a trade-off between low energy consumption and high quality of control. Fortunately, the EEAFS scheme performs well for different $β$ values in a very large range, e.g. from 1 to ∞, as shown in Figures 9 and 10. Consequently, it is easy to choose an appropriate $β$ in the settings of the EEAFS.

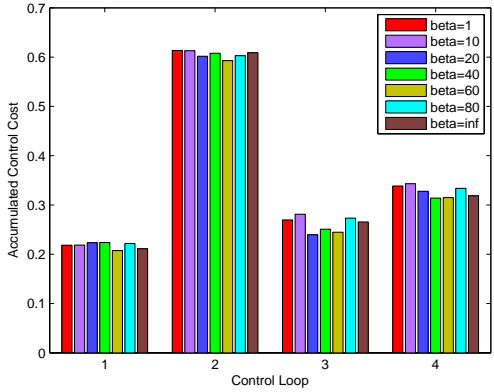

Figure 9. Accumulated loop control cost with different parameters

### C. Different Perturbation Intervals

The performance of the EEAFS in saving energy depends, to some degree, on the perturbations posed on each control loop, since task periods are adapted to control errors that are highly related to the perturbations. This set of simulations study this effect. All loops start running from t = 0, and are perturbed by step change input at the same time. The perturbation interval (PI) is set to be 1, 2, 4, and 6s, respectively, for each run. The whole simulation lasts 12s every time.

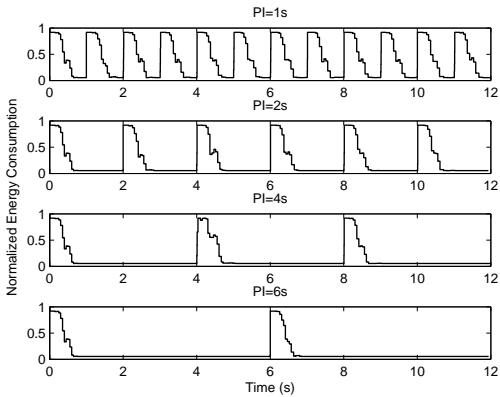

Figure 10. Energy consumption for different perturbation intervals

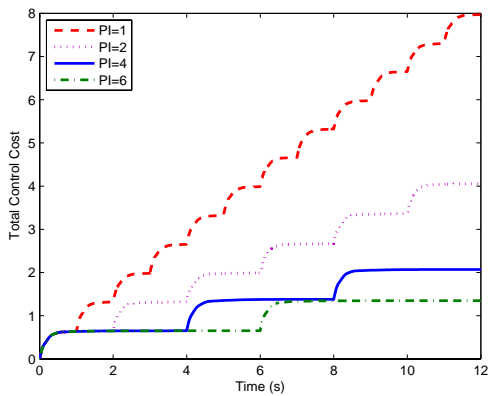

Figure 11. Total control cost for different perturbation intervals

With $\beta = 40$, Figures 10 and 11 give the CPU energy consumption and total control cost of the system for different perturbation intervals. Table II summarizes the average energy consumption and the total control cost under both opDVS and EEAFS.

TABLE II.
AVERAGE ENERGY CONSUMPTION AND TOTAL CONTROL COST

| PI (s) | $E_{AVG}$ | | | $J_{SUM}$ | | |
|---|---|---|---|---|---|---|
| | opDVS | EEAFS | Decrease | opDVS | EEAFS | Increase |
| 1 | 91.8% | 48.0% | 43.8% | 7.591 | 7.978 | 5.1% |
| 2 | 91.8% | 26.0% | 65.8% | 3.874 | 4.065 | 4.9% |
| 4 | 91.8% | 15.9% | 75.9% | 1.952 | 2.075 | 6.3% |
| 6 | 91.8% | 11.7% | 80.1% | 1.302 | 1.357 | 4.2% |

The CPU energy consumption under the opDVS remains constant regardless of changes in perturbation intervals. This is mainly due to the fact that the system workload remains changeless during run time, though the frequency of the perturbations varies.

When the EEAFS is employed, the CPU energy consumption becomes highly related to how often control loops encounter perturbations. It is seen that larger perturbation intervals yield lower energy expenditure under EEAFS. When PI = 1s, the average energy consumption under EEAFS is 48.0%, which is 43.8% less than under the opDVS. As PI increases, the energy consumption under EEAFS decreases accordingly and as a consequence, the additional energy consumption reduction over opDVS increases. When PI is set to 6s, the average energy consumption under EEAFS reduces to 11.7%, which is 80.1% less than that of the opDVS. The advantages of EEAFS become more paramount as the perturbations posed on control loops become lighter, i.e., when the frequency of perturbations is smaller.

The degradation of QoC under EEAFS over the opDVS is always minor for different perturbation intervals. The relative increase of the total control cost remains within 7%. This indicates that the EEAFS and the opDVS deliver comparable control performance.

## VI. CONCLUSION

By exploiting application adaptation, a feedback scheduling scheme has been proposed that can achieve additional energy saving over conventional pure DVS algorithms. In addition to CPU speed adjustment, the period of each control task is also adapted to relevant control performance, thus allocating computing resources among control tasks based on their real demands. Since task periods are enlarged provided that the QoC is not jeopardized, higher energy efficiency is achieved. Extensive simulation experiments have demonstrated the effectiveness of the proposed approach.

## ACKNOWLEDGMENT

This work is supported in part by China Postdoctoral Science Foundation under Grant No. 20070420232, Natural Science Foundation of China under Grant No. 60474064, and Zhejiang Provincial Natural Science Foundation of China under Grant No. Y107476 and No. Y1080685.

**Feng Xia** received the B.E. and Ph.D. degrees in information science and engineering (control discipline) from Zhejiang University, China, in 2001 and 2006, respectively. He was a visiting research fellow at Queensland University of Technology, Australia, from 2007 to 2008. He is currently an Assistant Professor at Zhejiang University. Dr. Xia serves on a number of Editorial Boards and Program Committees. In 2005 he was awarded Zhu Ke-Zhen Scholarship at Zhejiang University. He was also a recipient of the 2005 HP (Hewlett-Packard) Scholarship. He has published one book and over 50 refereed papers. His research interests include cyber-physical systems, wireless sensor/actuator networks, real-time and embedded systems, ambient intelligence, and real-time control. He is a member of IEEE and ACM.

**Longhua Ma** received his PhD degree from Zhejiang University, China, in 2001. He is currently an Associate Professor in College of Information Science and Engineering, Zhejiang University. His research interest covers modeling and optimization of complex systems, intelligent data processing, embedded computing systems, and real-time control.

**Wenhong Zhao** is a Professor in College of Mechanical and Electrical Engineering, Zhejiang University of Technology, China. His research interest covers embedded systems, real-time control, and electro-mechanical systems.

**Youxian Sun** is a Professor in College of Information Science and Engineering, Zhejiang University, China. He was a Visiting Scholar (Alexander Von Humboldt Fellow) at University of Stuttgart, Germany, from 1984 to 1987. His research fields include industrial automation, control theory and applications, etc. He is a Member of Chinese Academy of Engineering.

**Jinxiang Dong** is a Professor in College of Computer Science and Technology, Zhejiang University, China. He was a Visiting Professor at University of Southern California from 1993 to 1994. His research interests include computer graphics, artificial intelligence, database, information processing, security, advanced manufacturing and automation, etc.